\documentclass{ws-p8-50x6-00}

\usepackage{amssymb} 

\newcommand{\qpar}{q_\parallel^2}
\newcommand{\qperp}{q_\perp^2}

\title{
{\normalsize\rm E-\lowercase{print hep-ph/0001210} \hfill 
P\lowercase{reprint} YARU-HE-00/02} \\[5mm]
Mass Shift of Axion in Magnetic Field}

\author{N.V.~Mikheev, A.Ya.~Parkhomenko, and L.A.~Vassilevskaya} 

\address{Department of Theoretical Physics, 
         Yaroslavl State (Demidov) University, \\
         Sovietskaya 14, 150000 Yaroslavl, Russia \\ 
         E-mail: \, mikheev@yars.free.net \, parkh@uniyar.ac.ru \, 
                    lyuba@mail.desy.de}

\begin{document}

\maketitle

\abstracts{
A mass-shift of the axion propagating in an external constant homogenious 
magnetic field is calculated. The contributions via an electron 
loop and a virtual photon are examined. It is shown that the virtual 
photon contribution dominates substantially over the electron-loop one. 
Under the conditions of the early Universe the electron-loop contribution 
to the massless axion mass-shift is equal to zero while the virtual 
photon contribution is finite and can be of order of the recent 
restrictions on the axion mass.
} 

The axion~\cite{Peccei77,WW} obtains the mass due to the mixing 
with $\pi^0$-meson and, as a consequence, the Peccei-Quinn scale, 
$f_a$, is related to the axion mass, $m_a$, by the 
relation~\cite{Raffelt-book}: $m_a \simeq m_\pi f_\pi / f_a$, 
where $m_\pi$ and $f_\pi$ are the mass and the decay constant of 
$\pi^0$-meson. 
In models of an ``invisible'' axion~\cite{DFSZ,KSVZ} the axion mass 
is, in principle, arbitrary, however astrophysical and cosmological 
considerations provide an upper and lower bounds~\cite{Raffelt-book}: 
\begin{equation}
10^{-6}~{\rm eV} \lesssim m_a \lesssim 10^{-3}~{\rm eV} . 
\label{eq:ax-mass}
\end{equation}

The processes with weakly interacting particles, and with axions, 
in particular, are of importance under extreme external conditions 
which can be realized, in the early Universe as well as in astrophysical 
objects such as a magnetized neutron star or a supernova explosion. 
In studying of processes under such conditions one has to take into 
account non-trivial dispersions of particles. In considering axion 
processes the changing of the axion dispersion can occur substantial 
and, hence, should be investigated. In addition to the contribution 
to the axion self-energy via the electron loop, the other contribution 
via a virtual photon exists due to an effective axion-photon interaction 
in the external electromagnetic field. In this paper we show the 
importance of the photon-induced mass-shift of the axion in the strong 
magnetic field.   

The contribution to the axion mass squared,~$\delta m_a^2$, 
is connected with the real part of the field-induced 
amplitude $\Delta M$ of $a \to a$ transition by the relation:
\begin{eqnarray}
\delta m_a^2 = - \mbox{Re} \, \Delta M.
\label{eq:axmass}
\end{eqnarray}
%

%
\begin{figure}[tb]
%
\centerline{\epsfxsize=.4\textwidth \epsffile[175 580 335 670]{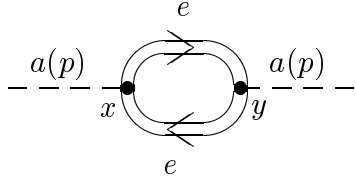}}
\caption{The external-field diagram of $a \to a$ transition via an 
         electron loop.}
\label{fig:polopaxion}
\end{figure}
%
%
In the second order of the perturbation theory $a \to a$ transition 
amplitude induced by the electron loop is described by the diagram 
shown in Fig.~\ref{fig:polopaxion}. Taking into account the external 
field influence means that one has to use exact electron propagators in the  
field which are drawn as double solid lines in Fig.~\ref{fig:polopaxion}. 
Below we consider the limit of the strong magnetic field when the field 
strength is the largest scale parameter ($|e B| \gg \qperp, \qpar, m_e^2$, 
where~$m_e$ and~$e$ are the mass and electric charge of the electron, 
$q_\mu$ is an axion four-momentum, $\qpar = q^2 + \qperp$,\footnote{
In calculations we use the metrics $g_{\mu \nu} = diag (1, -1, -1, -1)$,   
so that for any four-vector $a^2 = a_0^2 - {\bf a}^2$.} and $\qperp$ is 
the squared axion momentum component orthogonal to the magnetic field 
strength~{\bf B}). In this limit the electron-loop 
contribution,~$\left ( \delta m_a^2 \right )_e$, has a form~\cite{VMP-YaF}: 
\begin{equation}
\left ( \delta m_a^2 \right )_e \simeq 
       - \frac{g^2_{ae} \, |e B|}{2 \pi^2} \, 
       \exp \left ( - \frac{\qperp}{2 |e B|} \right ) \, 
       F \left ( \frac{4 m_e^2}{\qpar} \right ) . 
\label{eq:FIC-strong1} 
\end{equation}                               
Here, $g_{ae} = C_e m_e / f_a$ is an axion-electron coupling,
$C_e$ is a model-dependent parameter of order of unity, 
and the function $F (z)$ is: 
\begin{eqnarray}
F (z) =  
\left \{ 
\begin{array}{ll} 
\! \frac{\mbox{1}}{\mbox{$2 \sqrt{1 - z}$}} 
   \left [ 
   \ln \left | \frac{\mbox{$\sqrt{1 - z} - 1$}}
                    {\mbox{$\sqrt{1 - z} + 1$}} \right | 
   - i \pi \Theta (z) \Theta (1 - z) 
   \right ] , 
   & z < 1, \\[2mm] 
\! \frac{\mbox{\large 1}}{\mbox{$\sqrt{z - 1}$}} \, \arctan 
     \frac{\mbox{1}}{\mbox{$\sqrt{z - 1}$}} ,  
     & z \ge 1,
\end{array}
\right . 
\label{eq:F-our} 
\end{eqnarray}
where $\Theta (z)$ is the unit step-function. 
The imaginary part of~$F (z)$ in the kinematical region 
$0 < z < 1$ means that in the magnetic field the axion decay 
into an electron-positron pair $a \to e^+ e^-$~\cite{MV-PLB} 
is allowed under the condition $\qpar > 4 m_e^2$. 

\begin{figure}[tb]
\centerline{\epsfxsize=.4\textwidth \epsffile[160 500 330 550]{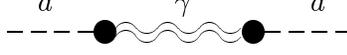}}
\caption{The external-field diagram of $a \to a$ transition 
         via a virtual photon.}
\label{fig:aga-loop}
\end{figure}

The other contribution to the axion mass squared in the external 
field via the virtual photon is described by the diagram shown in 
Fig~\ref{fig:aga-loop}. 
%
%
%
%
We note that it is necessary to take into account the influence of the 
field on both the axion-photon coupling and the photon propagator. 
The photon contribution to the field-induced axion mass squared has 
a form~\cite{VMP-YaF}:
\begin{equation}
\left ( \delta m_a^2 \right )_\gamma = 
\bar g^2_{a\gamma} B^2 \qpar \, {\rm Re}  
\left ( q^2 - \mbox{\ae}^{(2)} \right )^{-1} . 
\label{eq:FIC-strong3} 
\end{equation}
Here, $\bar g_{a \gamma} = \alpha \bar \xi / 2 \pi f_a$ is an effective 
axion-photon coupling, $\alpha$ is the fine-structure constant, 
$\bar \xi$ is a model-dependent parameter of order of unity 
with taking into account the external field corrections~\cite{MRV-PRD}, 
and \ae${}^{(2)}$ is an eigenvalue of the polarization operator 
corresponding the ``second'' photon mode in the notation of Ref.~6:  
\begin{equation}
\mbox{\ae}^{(2)} \simeq - \frac{2 \alpha |e B|}{\pi} \, 
\left [ \frac{4 m_e^2}{\qpar} 
F \left ( \frac{4 m^2}{\qpar} \right ) - 1 \right ] ,
\label{eq:kappa2-smf} 
\end{equation}
where the function $F (z)$ is defined in Eq.~(\ref{eq:F-our}). 

The numerical estimations of the field-induced contributions 
to the axion mass-shift (the self-energy in the axion rest frame) 
are~\cite{VMP-YaF}: 
\begin{eqnarray}
(\delta m_a^2)_e & \simeq & 
      - 3.3 \times 10^{-31} \, {\rm eV}^2 \times
        C_e^2 \, \frac{B}{B_0} \,
        \left ( \frac{10^8 \, {\rm GeV}}{f_a} \right )^2 
        \left ( \frac{m_a}{10^{-3} \, {\rm eV}} \right )^2,  
\label{eq:ff-shift-estim} \\
(\delta m_a^2)_\gamma & \simeq & 1.26 \times 10^{-15} \, {\rm eV}^2 \times 
        \xi^2 \left ( \frac{B}{B_0} \right )^2 
        \left ( \frac{10^8 \, {\rm GeV}}{f_a} \right )^2 , 
\label{eq:ph-shift-estim1} 
\end{eqnarray}
where $B_0 = m_e^2 / |e| = 4.41 \times 10^{13}$~G is the so-called 
Schwinger value. For the magnetic fields strength of order of the 
Schwinger value the long-range contribution via the virtual photon 
is $10^{16}$ times larger then the electron-loop one but both 
contributions are very small. It means that in studying axion processes 
in magnetized astrophysical objects with the field strength 
$B \sim 10^{13} - 10^{15}$~G one can neglect the field influence 
on the axion mass~(\ref{eq:ax-mass}). 

However the situation is possible when the field-induced contribution  
to the axion mass-shift is essential. It is the case when the axion 
is a massless particle as, for example, before the QCD phase transition 
under the conditions of the early Universe. Note that at this stage of 
the Universe evolution a very strong magnetic field with the strength 
$B \sim 10^{22} - 10^{23}$~G can exist~\cite{Universe}. 
In this case the electron-loop contribution to the axion self-energy, 
$(\delta m_a^2)_e$, which is proportional to the axion transfer momentum, 
is equal to zero in the axion rest frame while the virtual photon one 
is not vanish~\cite{VMP-YaF}:   
\begin{equation}
\delta m_a \simeq 0.058 \, {\rm eV} \times \xi 
        \left ( \frac{10^8 \, {\rm GeV}}{f_a} \right )   
        \left ( \frac{B}{10^{23} \, {\rm G}} \right )^{1/2} .  
\label{eq:ph-shift-estim2}  
\end{equation}
It is seen that under the early Universe conditions the massless axion 
acquires a mass due to the long-range interaction with photons.  
As the model-dependent parameter~$\xi$ is typically of order of 
unity~\cite{Raffelt-book}, the axion mass induced by a primordial magnetic 
field can be as large as its recent value~(\ref{eq:ax-mass}).  

\section*{Acknowledgments} 

This work was partially supported by INTAS under grant No.~96-0659 
and by the Russian Foundation for Basic Research under 
grant No.~98-02-16694.

\end{document}